Systematic identification of gene families for use as "markers" for phylogenetic and phylogeny-driven ecological studies of bacteria and archaea and their major subgroups

Dongying Wu[1,2*], Guillaume Jospin[1], Jonathan A. Eisen[1]

1. U. C. Davis Genome Center, University of California Davis, Davis, California 95616, USA
2. DOE Joint Genome Institute, Walnut Creek, California 94598, USA

* Corresponding author

## Abstract

With the astonishing rate that the genomic and metagenomic sequence data sets are accumulating, there are many reasons to constrain the data analyses. One approach to such constrained analyses is to focus on select subsets of gene families that are particularly well suited for the tasks at hand. Such gene families have generally been referred to as "marker" genes. We are particularly interested in identifying and using such marker genes for phylogenetic and phylogeny-driven ecological studies of microbes and their communities (e.g., construction of species trees, phylogenetic based assignment of metagenomic sequence reads to taxonomic groups, phylogeny-based assessment of alpha- and beta-diversity of microbial communities from metagenomic data). We therefore refer to these as PhyEco (for phylogenetic and phylogenetic ecology) markers. The dual use of these PhyEco markers means that we needed to develop and apply a set of somewhat novel criteria for identification of the best candidates for such markers. The criteria we focused on included universality across the taxa of interest, ability to be used to produce robust phylogenetic trees that reflect as much as possible the evolution of the species from which the genes come, and low variation in copy number across taxa.

We describe here an automated protocol for identifying potential PhyEco markers from a set of complete genome sequences. The protocol combines rapid searching, clustering and phylogenetic tree building algorithms to generate protein families that meet the criteria listed above. We report here the identification of PhyEco markers for different taxonomic levels including 40 for "all bacteria and archaea", 114 for "all bacteria (greatly expanding on the ~30 commonly used), and 100s to 1000s for some of the individual phyla of bacteria. This new list of PhyEco markers should allow much more detailed automated phylogenetic and phylogenetic ecology analyses of these groups than possible previously.

Introduction

In the 1970s, pioneering studies were carried out by Carl Woese and colleagues to analyze the sequences of fragments of rRNA genes [1-3]. They demonstrated that by analyzing the sequence information found in the small subunit rRNA (ssu-rRNA), we can place the diverse cellular organisms in a tree of life [2,3]. The ssu-rRNA also led Woese et. al. to discover Archaea, the third domain of life in addition to the already known Bacteria and Eukaryotes [2,3]. Since then, ssu-rRNA has been widely adopted as a "*phylogenetic marker*" for studies of diverse living organisms. The ssu-rRNA gene has many advantages to play such a role. It is present universally in all cellular organisms across all three domains of life. The sequences of ssu-rRNA have desirable patterns such as diverse regions separated by highly conserved regions. The conservation at the sequence and structure level facilitate the studies that require sequence alignments [4,5], while sequence variations provide valuable information for analysis of both recent and ancient evolutionary events [5,6].

Since (and even before) the time of Woese's work on rRNA, other genes have been developed into widely used and robust phylogenetic markers for various taxa. However for global studies of microorganisms, ssu-rRNA genes are still the phylogenetic marker of choice [6]. A related topic to general studies of the phylogeny of microbes is that for a long time, scientists only studied cultured microbial organisms and left out the overwhelming majority that could not be cultured in the lab [7]. A fundamental shift occurred when researchers started examining rRNA genes from organisms never grown in the lab. This work accelerated in particular when the polymerase chain reaction (PCR) methodology was adapted in ssu-rRNA studies [8]. The highly conserved regions of the ssu-rRNA allow one to design oligonucleotide primers to amplify the ssu-rRNA genes by PCR for a diverse range of species. Using "universal primers" in PCR, scientists can amplify ssu-rRNA from a wide variety of taxa directly from the environment in a single reaction [8,9]. The culture independent PCR amplification and sequencing of ssu-rRNA from an unprecedented variety of communities is enriching ssu-rRNA sequence collections exponentially. By the start of 2012, the SILVA ssu-rRNA database has reached 2,492,653 sequences [10], while the RDP database has included 2,110,258 ssu-rRNA sequences [11]. With next generation sequencing there are also billions if not trillions of available partial ssu-rRNA sequences. Thus in addition to its use as a phylogenetic marker, ssu-rRNA became a key "ecological marker" for studies of microbes.

Despite the power and ongoing potential of ssu-rRNA based studies of microbial diversity, using this gene – or just this gene – has its limitations. For example, the extensive variation in copy number of ssu-rRNA genes among different organisms poses major challenges for using ssu-rRNA as an ecological marker [12-14]. This is because researchers try to use counts of the number of sequences retrieved from a particular group to estimate relative abundance of such groups. Although methods have been developed to try and correct for the variance in rRNA copy number [15], but they are imperfect. Another limitation of ssu-rRNA only studies is that all "universal primers" for PCR amplification of ssu-RNA genes have different degrees of bias, they usually prefer certain taxonomic groups over others [14]. In addition, phylogenetic trees built from one gene do not always reflect the true evolution history of species under study. Any gene, including ssu-rRNA, is subject to horizontal gene transfer, convergent evolution, or evolution rate variations between different phylogenetic groups [16-18] and other forces that can lead to the trees of that gene not accurately reflecting the history of species.

Over the years researchers have attempted to identify, develop and use other marker genes to for microbial diversity studies to compensate for the limitations of ssu-RNA genes as markers. One example is the recombinase A gene family that includes bacterial RecA [19,20], archaea

RadA and RadB [21], eukaryotic Rad51 and Rad57[22], phage UvsX [23]. The genes in the recA superfamily are crucial for recombination and DNA repair, are nearly universally present, and have (compared to ssu-rRNA) low variation in copy number between taxa. Another example is RNA polymerase β subunit (RpoB) gene, which is responsible for transcription initiation and elongation [24-26]. Both genes have been widely used in phylogenetic studies of bacteria, archaea, and eukaryota [27,28]. In terms of use as phylogenetic markers, protein-coding genes have some advantages over ssu-rRNA including that they may have less (or at least different) nucleotide compositional bias than ssu-rRNAs [29,30].

There are two major challenges for using protein-coding genes as phylogenetic markers for broad studies of microbial diversity. First, PCR amplification technology, which is the driving force behind using ssu-rRNA for microbial diversity studies, does not work as easily for protein coding genes. This is because DNA level variations can be observed even for highly conserved protein domains at the amino acid level. On one hand, such DNA level variations are valuable for the studies of closely related organisms [30]; on the other hand, the primers for PCR amplification of these genes need to be degenerate, sometimes extremely so [21]. Thus phylogenetic analysis of protein-coding genes has largely focused on cultured organisms because of the limited ability to sequence such genes from unknown organisms using PCR-based methods. Metagenomics, the direct sequencing of the organisms present in the environmental samples without PCR, has made protein based phylogenetic analysis and phylogeny-driven ecological analysis of uncultured organisms feasible [27,28].

In general, metagenomics is opening up the possibility that any gene can be used for studies of microbial diversity. However, one challenge in this is the lack of knowledge about what genes are suitable for such studies. For broad studies of microbial diversity (e.g., studies of the diversity of all bacteria in metagenomic data or whole-genome phylogenetics of all bacteria) the most widely used sets of genes include about 30 genes (e.g., 31 in [30] and [31]).

Though the previously identified marker gene sets are useful, we became interested in revisiting marker gene identification and in developing and using a system that would be updated in multiple areas. Some of the key limitations in the previously identified marker sets included that they were selected when only a small number of genomes were available, the methods behind their identification were not fully automated, the sets were focused on the highest level taxonomic groups and thus missed genes that could be useful for more narrow focus on specific subgroups of bacteria or archaea and the sets were focused largely on markers for phylogenetic studies not for phylogeny-driven ecological studies.

We report here the development of an automated approach to identify phylogenetic and phylogenetic ecology markers (and thus refer to these as PhyEco markers). Our approach takes a set of complete genome sequences and applies a variety of criteria for assessing the gene families present in those genomes for their potential use as PhyEco markers. The criteria we use includes universality across the taxa of interest (which is important for multiple reasons), ability to be used to produce robust phylogenetic trees that reflect as much as possible the evolution of the species from which the genes come (which helps control for issues like convergent evolution and lateral gene transfer), and low variation in copy number across taxa (which allows for markers to be used for estimates of relative abundance of taxa). The protocol takes all the proteins in all the genomes under consideration and, using rapid searching and clustering algorithms, generate protein families from that complete protein set. Phylogenetic trees are then built for each family and subgroups in the trees (i.e., clades) are automatically sampled and evaluated for the criteria listed above. Potential PhyEco marker families are then further assessed using multiple comparative and phylogenetic analyses.

Our systematic approach reveals 40 PhyEco marker candidates spanning the domains of bacteria and archaea. Our analysis also identified 74 bacterial specific PhyEco markers, which, with the 40 bacterial and archaeal markers, brings the total to 114 PhyEco markers that can be used for analysis of bacteria. In addition, our analysis revealed 100s – 1000s of phyla-specific PhyEco marker genes. After we finished the work described here and presented it at multiple meetings (but before we submitted it) a paper was published from Wang and Wu [32] describing a similar approach to identify taxa specific phylogenetic markers. We note one of the authors of that paper (Martin Wu) was working in the Eisen lab when we started this analysis. We could get into all sorts of discussions of the complex history here but we do not believe that would be particularly useful. Suffice it to say we were surprised to see this paper from his lab. Regardless of the history, our methods are different than those used by Wang and Wu and are results also have differences. We therefore present our work here as an independent development, which is what it is. Perhaps most importantly, the new list of PhyEco markers we have identified should allow much more detailed automated phylogenetic and phylogenetic ecology analyses of these groups than possible previously.

Results and Discussion

**De novo identification of protein families from massive and ever increasing genome data sets by a bottom up approach**

For our study here we decided that it was important to first perform a de novo build of protein families from currently available genome data sets (i.e. we did not want to rely on existing protein family information). Largely we chose this approach because of concerns about the possibility of bias in the existing protein family data sets related to their being created and built when the sampling of genomes was phylogenetically very limited. To carry out a de novo identification of protein families we needed to search all the proteins from all the genomes of interest against each other (e.g., by using an all vs. all BLAST [33] search). Then we would need to group these proteins into families based on the results of such searches. Given the large number of genomes available for bacteria and archaea, such all vs. all searching and then clustering would be computationally very intensive. Furthermore, we wanted to be able to do such a de novo build again and again in some sort of automated manner as more genomes became available. Such an accelerating and increasing costly approach was not possible within the scope of this project. The computational infrastructure we had available for this work could comfortably handle de novo gene family building and analysis for ~200 bacterial and/or archaeal genomes at a time.

The limitation outlined above led us to develop a "bottom up" strategy for identifying PhyEco markers. This strategy worked in the following general way: the total set of genomes was divided into subgroups based on phylogeny and taxonomy of the species from which the genomes came. In most cases subgroups corresponded to phyla. However, there were too many genomes for some phyla (e.g., proteobacteria) and thus we further subdivided the group (in the case of the proteobacteria we divided it into the five major classes – alpha, beta, gamma, delta and epsilon). Then for each of these subgroups we carried out all vs. all searches using BLASTP [33] of the proteins encoded in the genomes. Following this, for each subgroup we used the results of the all vs. all search to create protein families using the MCL clustering algorithm [34].

The use of MCL for gene family building has some advantages and disadvantages. The main advantages relate to speed and ability to control the granularity of the output clustering [34].

However this comes with a risk of splitting up what should be single families into separate clusters [34]. To compensate (at least somewhat) for this risk of splitting, after the MCL clustering is run, the left over sequences are clustered using a more aggressive single linkage clustering method. This allows the recovery of some additional protein families that were artificially split by the MCL method.

**Identifying PhyEco markers from protein family sets**

To identify PhyEco markers we screened through the complete protein family data sets using four criteria: universality, evenness in copy number, monophyly and uniqueness. We selected these specific metrics for multiple reasons. Universality (how widely found the gene family is for a group in question) is important for phylogenetic studies because the more universal (for a particular group) a gene is, the less missing data one would have in a data set for phylogenetic analysis. For ecological studies, universality is important because it allows one to assume that if a representative of that taxonomic group were present it would be likely to have that particular gene. Evenness in copy number is important in particular in the estimation of the relative abundance of taxa in environments. Regarding "monophyly", what we wanted to do was develop a metric that would allow us to identify those cases where the genes from the taxonomic group in question were monophyletic in the phylogenetic tree for the whole family. For example consider a hypothetical protein family "A" being screened PhyEco marker potential for the Cyanobacteria. We wanted to know, in a phylogenetic tree including all homologs of protein family A, if the proteins from Cyanobacteria all grouped together as a single clade. In order to identify PhyEco markers that captured our current understanding of phylogenetic/taxonomic structure of bacteria and archaea, we examine phylogenetic trees to make sure they are monophyletic for the phyla and classes that are regarded as monophyletic based on previous phylogenetic studies (e.g., [35]). Finally, we developed a "uniqueness" test as a way to measure how well we were able to distinguish members of the PhyEco marker family of interest from other families. This test is based on how we currently search for members of a particular family in a genomic or metagenomic data set. We do this by creating a hidden markov model (HMM) representing a sequence alignment of the family, and then we search the data sets of interest for sequences in the family using the HMM. A family's HMM (and thus the family) is determined to be unique if a HMM search against all families retrieves all the target family members as top hits with a comfortable distance from sequences in other families.

For identification of PhyEco markers for each of the subgroups we used these metrics in a mostly qualitative manner. For a protein family to be included in the list for a group, it had to (1) be present in all members of the group (universal) in a single copy (even). This was a very strict set of criteria to use. In the future we will likely have to relax this approach a bit as more and more genomes that are incomplete become available but could be useful to include [36].

After PhyEco marker families have been identified for each subgroup, we then coalesce them together to identify PhyEco markers for the "higher" taxonomic levels. For example the families in the different classes of proteobacteria were compared to identify proteobacteria-wide PhyEco marker families. In addition, the PhyEco markers for each of the bacterial phyla were compared to identify bacterial-wide PhyEco markers. The comparison across subgroups is done in the following way. Consensus sequences are generated for each PhyEco marker family for each subgroup. These are then all compared to each other and clustered using single linkage clustering to identify which subgroup specific PhyEco marker families are related to each other (see Methods for more detail). Once we coalesced together the families from the lower levels we then scored candidate PhyEco families for the "higher" levels using measures of universality, evenness and monophyly with quantitative metrics for each as described in the Methods.

We are experimenting with another strategy to avoid large-scale de novo gene family building in the future. We have been involved in creating a database of gene families for all the sequenced genomes called SFAMs with an updating protocol that adds future sequences into existing gene families and only builds novel gene families if necessary [37]. In theory, this database could be used directly and the families could be screened for criteria of interest (such as universality, evenness, uniqueness, and monophyly) for any taxonomic group of interest.

**A diverse sampling of a taxonomic group is essential for PhyEco marker identification**

For the analyses reported here, we focused only on phylogenetic groups for which a large number of complete genome sequences were available. Specifically, we considered the following taxonomic groups: "all bacteria and archaea", "all bacteria", "all archaea", each phylum of bacteria for which there were more than a few genomes (there were generally not enough genomes for specific archaeal phyla to subdivide the archaea up), and the classes within the proteobacteria phylum (there are many genomes for each class due to biases in genome sequencing efforts) [35].

We have identified taxonomic group-specific PhyEco marker candidates systematically for 17 taxonomic groups (table 1). The number of candidate PhyEco markers varies for different taxonomic groups: e.g., the Deinococcus-Thermus phylum has 974 PhyEco markers, while Firmicutes has only 87. Because of the low number of genomes and low phylogenetic diversity at the phyla level, Deinococcus-Thermus, Thermotogae and Chlamydiae have exceptionally high number of PhyEco markers (the fewer genomes the less likely there is to have something unusual in one of the genomes that would remove a gene from our list – and the lower the phylogenetic diversity the more genes there will be that are shared). Taxonomic groups with more genomes and higher phylogenetic diversity tend to have fewer PhyEco marker genes using our criteria. The most diverse groups have about 100, they can be regarded as "core" genes that are relatively resistant to gene duplication, deletions and transfer [38-40]. We have identified 560 PhyEco markers for Cyanobacteria, an extraordinary large number for such a phylogenetic diverse group. The presence of many photosynthesis related genes contributes to the large number [40]. Certain lineages, such as $\gamma$-proteobacteria, have such large numbers of genomes from a limited number of species that we have to use phylogenetic trees to select and effectively reduce the number of genomes to facilitate de novo gene family building. But for most phyla of bacteria and archaea, more genome sequences are in order to more accurately identify good PhyEco markers. Efforts such as the phylogeny-driven Genomic Encyclopedia of Bacteria and Archaea (GEBA) project which focuses on cultured organisms [35] and culture independent genome sampling efforts such as [41,42] would be very beneficial to continue in relation to PhyEco marker identification.

**Properties of the "All Bacteria and Archaea" PhyEco marker set**

Our analysis identified 40 PhyEco markers for the group "all bacteria and archaea" (Table 1). We have assessed these markers in a few ways. First, examination of phylogenetic trees of each of these PhyEco markers shows that for each, Archaea and Bacteria form two distinctive clades (Figure 1). In addition, we examined the properties of each of these markers within the lower taxonomic groups in our data set. For example, one of the "all Bacteria and Archaea" PhyEco markers corresponds to ribosomal protein S2. For all 18 of the taxonomic groups examined in our analysis (e.g., all Bacteria, all Archaea, and the major bacterial subgroups) this protein also showed up as PhyEco marker. All organisms have a single copy of ribosomal protein S2 gene, and each of the 18 taxonomic group forms a monophyletic clade in the family phylogenetic tree. All the other 39 "all bacteria and archaea" PhyEco markers were identified as PhyEco markers at a minimum of seven taxonomic levels (Figure 1). These results are not overly surprising since we inferred the "all bacteria and archaea" markers by building up from the lower levels. But in theory it is possible that something would not work in this approach. For our "building up" method might not work as we expected. In addition, there might be more conflicts that we expect between taxonomy (which is largely on ssu-rRNA phylogeny) and protein family phylogeny. Another key feature of the "all bacteria and archaea" PhyEco markers is their functional roles in organisms. The vast majority of them are associated with the translation processes (Table 1): 30 ribosomal protein subunit genes, 1 translation initial factor, 1 translation elongation factor, and 3 rRNA synthesis related genes. The rest of are involved in protein metabolism including peptide degradation and exporting, RNA degradation, heme biosynthesis and purine nucleotide synthesis.

**Properties of the "All Bacteria" PhyEco marker set**

Our analysis identified 74 PhyEco markers for the group "all bacteria." As with the "all bacteria and archaea" PhyEco markers we assessed these in a variety of ways. As expected, all 74 have very low variance in copy number between bacterial genomes and almost all are present in only a single copy in all bacterial genomes. Interestingly, many have no obvious counterpart in archaea (Figure 2). In terms of lower taxonomic groups (in this case – the subgroups within the bacteria), each of the 74 "all bacteria" PhyEco markers was also identified as a PhyEco marker for at least six taxonomic groups (Figure 2). Similar to the "all archaea and bacteria" markers, translation related gene families dominate the list of "all bacteria" markers (Table 1S): fourteen are ribosomal protein genes; ten are related to tRNA synthesis and modification; three are involved in translation initiation, elongation and termination; others are for ribosome rescue and recycle, and small subunit rRNA processing. DNA replication and repair related gene families are abundant in the list: four families are related to DNA synthesis and twelve are involved in DNA repair. Others of these bacterial PhyEco marker gene families are involved in transcription, RNA degradation, protein trafficking and degradation, cell division and shaping and signal transduction. Various biosynthesis related families are also included in the PhyEco markers, including the synthesis of cofactors, peptidoglycan, pyrimidine, fatty acid and heme. Interestingly, three of the "all bacteria" PhyEco marker candidates have no functional annotation and apparently have not been studied functionally even though their distribution patterns among the bacterial genomes strongly suggest that they are potentially involved in essential processes.

The automated pipeline for phylogenomic analysis AMPHORA (co-developed by one of us) included 31 phylogenetic markers for bacteria according to previous experiments [30]. We are glad to see 30 of the original 31 AMPHORA markers are included in our list of PhyEco markers: 18 are included in the "bacteria and archaea" PhyEco markers, while twelve are included the "all bacteria" PhyEco markers (Table 2, Table S1). The only AMPHORA marker missing from our list is the phosphoglycerate kinase gene (pgk). The phosphoglycerate kinase gene family is a good marker candidate at only 5 taxonomic levels (Bacteria, Chlamydiae, Cyanobacteria, Thermotogae, γ-proteobacteria), thus is not qualified to be one of the 114 core PhyEco markers of bacteria in this release.

**Conclusion**

The approach we describe here can be used for any taxonomic group and can also be automated for keeping up with the explosion in genome sequences that are coming. A core set of robust "PhyEco markers" has many uses and the metrics we describe here can help in objectively selecting candidates for such markers. We note that in the future it will likely be even better to go beyond PhyEco marker genes to incorporate information about additional gene families in analysis of genomes and metagenomes. One approach is to identify families for which the presence indicates a particular clade [43]. We have taken an alternative approach, that is to build resources and tools that will allow phylogenetic analysis of all families from samples. This is why we developed the SFAMs database [37], and also why we are helping develop phylosift which can make use of any family in automated phylogenetic analyses of metagenomic data.

**Availability**

The gene families of potential PhyEco markers listed in Table 1 and Table S1, as well as phyla-specific PhyEco markers, are available for download. The package includes hidden markov model profiles, amino acid sequences and alignments. The phylogenetic trees, perl scripts for calculating gene family universality, evenness and monophyletic values, as well as the numeric values of the three measurements for Figure 2 and 3 are also available and accessible from: http://edhar.genomecenter.ucdavis.edu/~dwu/BAmarker/.
We have also uploaded the files to Figshare, which can be accessed from: http://figshare.com/articles/Systematically_identify_phylogenetic_markers_at_different_taxonomic_levels_for_bacteria_and_archaea/722713

Methods

**Genome selection and database setup**

Bacterial and archaeal genomes from the IMG database were selected for phylogenetic marker identification [36]. Only genomes that were complete were included. We started our marker identification process at 15 different taxonomic levels: the domain Archaea; the phyla Actinobacteria, Bacteroides, Chlamydiae, Chloroflexi, Cyanobacteria, Firmicutes, Spirochaetes, Deinococcus-Thermus and Thermotogae; the classes α–proteobacteria, β-proteobacteria, γ–proteobacteria, δ -proteobacteria and ε-proteobacteria. Genomes that have undergone major genome reductions, such as those of Mycoplasma [44,45] and γ-proteobacteria endosymbionts [46,47], were not included in this study. Only one strain of the same species within γ-proteobacteria was selected if other strains did not contribute to the phylogenetic diversity (PD) in the phylogenetic tree of bacteria [35].

A phylogenetic tree was built for all the genomes in the selection. Alignments of ssu-rRNAs were extracted from the greengenes database[48]. Fasttree was used for ssu-rRNA tree building[49].

**Measurement of universality, evenness, monophyly and uniqueness for gene families for identifying PhyEco marker candidates**

To determine if a family is a suitable candidate of a PhyEco marker for a given taxonomic group, we developed four measurements: universality, evenness, monophyly, and uniqueness.

Universality - how widely distributed a family is across a taxonomic group. Universality is defined by the following equation:

$$U = 100 \times n/t$$

U is the universality value of a family, *t* is the total number of genomes in the taxonomic group of interest, *n* is the number of genomes in which the family can be found.

Evenness – how uniform the number of representatives per genome is for a family for a taxonomic group. Evenness is defined by the following equation:

$$E = 100 \times e^{\frac{-4 \times \sum |N_i - N_a|}{n}}$$

E is evenness value of a family, *Ni* is the number of family members from the genome *i*, *Na* is the average number of family members per genome, *n* is the number of genomes in which the family can be found.

Monophyly – for any family, how monophyletic are the representatives from a particular taxonomic group in a phylogenetic tree of the family across all genomes. We developed a "monophyletic value", which is based on Shannon entropy [50], to measure the topological distribution of a family member in a tree. If a family member can be broken into a number of monophyletic clades, the equation that defines monophyletic value is given as:

$$M = 100 \times e^{0.75 \times \sum [(c_i / n) \times \ln(c_i / n)]}$$

*M* is the monophyletic value, *n* is the total number of family members in the tree, $c_i$ is the number of family members in the monophyletic clade *i*.

We note our monophyly metric was designed to identify cases where the phylogenetic tree of a particular gene family is similar to the expected phylogenetic tree for the species being analyzed. However, the metric we used only reflects tree topology, and it is only effective in capturing almost perfect monophyletic clades. We are in the process of improving the monophyletic measurement to be more robust and reflect both tree topologies and branch lengths.

<u>Uniqueness</u> - how distinct is the family in question from other families. For this test we used profile hidden Markov models (profile HMMs)[51] of each family. Uniqueness was measured in terms of the results of searches of the family HMM profile against all the peptide sequences encoded in the genomes in the taxonomic group. The HMM search bit-score of any non-family-member sequence has to be lower than all the family members for a gene family to be considered unique. The larger the distances between the family-members and non-family-members, the more "unique" the family is. For the analysis reported in this paper we used the following approach:

$P_{member}$ is the worst hmmsearch P value of the family members, $P_{nonmember}$ is the best hmmsearch P value of non-family members, we only consider a family distinct if $P_{member}$ and $P_{nonmember}$ satisfy the following condition:

$$\lg E_{member} \leq \lg E_{nonmember} - 16.24 \times e^{-0.015 \times \lg E_{nonmember}}$$

**Identification of PhyEco markers for a taxonomic group**

All vs. all BLASTP searches were performed for the peptide sequences encoded in all the genomes in a taxonomic group of interest using an expected value cutoff of 1e-10 [33]. Those pairs of proteins for the BLASTP hits covered 80% of the query and hit sequences were considered "linked". The BLASTP similarity scores were retrieved for all links. The Markov Cluster Algorithm (MCL clustering) was performed for the links using an inflation value of two [34]. The resulting MCL clusters were regarded as families. The families with four to 2000 members were then used for PhyEco marker identification.

Each family was analyzed separately for potential as a PhyEco marker using the protocol illustrated in the flowchart in Figure 1. First, the peptide sequences from the family are aligned by MUSCLE [52] and then a phylogenetic tree is inferred from the alignment using Fasttree [49]. The subfamilies defined by the clades in the tree are then analyzed one at a time for universality and evenness. Those subfamilies that met the following criteria were considered for further analysis as PhyEco candidates: present in all or all but one of the genomes (universal) and present only once in all or all but one of the genomes (even). For each subfamily that passed the universality and evenness screening, the sequences were aligned by MUSCLE[52] and HMM models were built from the alignments by HMMER3[51]. The HMM for each subfamily was then searched against the entire collection of proteins for all the genomes in the phylogenetic group of interest [51]. Subfamily uniqueness was measured as described above with only those passing the selection criteria being identified as "unique." The final list of PhyEco markers for any group were thus those that passed the universality, evenness and uniqueness tests of this protocol.

As discussed above, the MCL method has the potential to mistakenly split up families that should be together. To attempt to correct for this limitation, we carried out single linkage clustering to build another round of gene families. First, BLASTP links that connected to members of PhyEco marker families were excluded. Second, gene families were built by using the single linkage clustering algorithm on the remaining BLASTP links. Finally, the gene families were subjected to the same protocol as outline above to identify additional PhyEco marker candidates.

**Identification of PhyEco markers for "higher groups" by coalescing together markers from "lower" groups**

In our bottom up approach, to identify PhyEco markers for the "higher" level taxonomic groups (e.g., all bacteria plus archaea) we needed to coalesce together PhyEco markers identified for each "lower" level phylogenetic group. This coalescing was done in the following way. First, one representative sequence was generated from each of the PhyEco markers' HMM models by hmmemit from HMMER3[51]. Then, an "All vs. All" BLASTP was performed for the representative sequences using an e-value cutoff of 1e-3 [33]. This was followed by single linkage clustering of the BLASTP results to generate clusters of lower level PhyEco markers that were similar to each other. We focused subsequent analysis on the 404 of these clusters that contained more than three of the lower level PhyEco marker representative sequences. For each of the 404 clusters, alignments were built with MUSCLE [52] followed by phylogenetic tree inference using Fasttree [49].

All clades in the trees that contain single representative from different taxonomic groups in the study were gathered, and the uniqueness of was measured using the approach outlined above. 382 "unique" clades were identified that each covered more than four taxonomic groups, and all the sequences they represented were retrieved to form 382 superfamilies. For each of these 382 superfamilies, a HMM profile was built and hmmsearch was performed against all complete bacterial and archaeal genomes [51]. The hmmsearch results were manually examined, and additional sequences identified by the search were retrieved and included in the superfamily. Alignments were built for all the superfamilies by MUSCLE [52] and phylogenetic trees were built by PHYML using JTT models [53]. Then, for each of the 382 superfamilies, universality, evenness and monophyletic values were calculated at each of the following 18 taxonomic levels: the domain Archaea and Bacteria; the phyla Actinobacteria, Bacteroides, Chlamydiae, Chloroflexi, Cyanobacteria, Firmicutes, Proteobacteria, Spirochaetes, Deinococcus-Thermus and Thermotogae; the super-class $\beta\gamma$-proteobacteria; the classes $\alpha$-proteobacteria, $\beta$-proteobacteria, $\gamma$-proteobacteria, $\delta$-proteobacteria and $\epsilon$-proteobacteria.

We then used these results to select PhyEco markers for the different "higher" taxonomic groups. To select PhyEco markers for the group "all bacteria and archaea" we required the product of universality, evenness and monophyletic values to be greater than 729,000 for at least a subset of taxonomic levels. We picked this value because it represents the value we would see with a score of 90 for each metric. We required this value to be exceeded in a minimum of seven of the taxonomic levels measured including the "all bacteria" and "all archaeal" sets. 40 families meeting this criterion were identified. To select PhyEco markers for the group "all bacteria" we again required the product of universality, evenness and monophyletic values to exceed 729,000 for a subset of the taxonomic levels included in this group. In this case we required this value to be exceeded in a minimum of six of the taxonomic levels measured (including the "all bacteria" one). 74 "all bacteria" PhyEco markers were identified with these restrictions.

Tables

Table 1: Summary of taxonomic group-specific PhyEco marker candidates. PD (phylogenetic distance) and monophyletic value is based on a PHYML tree of small subunit rRNA of all the 666 genomes in the study. The total PD of the ssu-rRNA tree is 82.70.

| Taxonomic group | Genome Number | Gene Number | PD Coverage | Monophyletic Value | PhyEco Marker Candidates |
|---|---|---|---|---|---|
| Bacteria and Archaea | 666 | 2,271,359 | 82.70 | NA | 40 |
| Archaea | 62 | 145,415 | 12.15 | 100.00 | 106 |
| Bacteria | 604 | 2,125,944 | 69.23 | 100.00 | 114 |
| Actinobacteria | 63 | 267,783 | 6.84 | 100.00 | 136 |
| Bacteroides | 25 | 71,531 | 5.12 | 100.00 | 286 |
| Chlamydiae | 13 | 13,823 | 0.69 | 100.00 | 560 |
| Chloroflexi | 10 | 33,577 | 2.66 | 100.00 | 323 |
| Cyanobacteria | 36 | 124,080 | 2.88 | 100.00 | 590 |
| Deinococcus-Thermus | 5 | 14,160 | 0.98 | 100.00 | 974 |
| Firmicutes | 106 | 312,309 | 13.49 | 88.70 | 87 |
| Spirochaetes | 18 | 38,832 | 2.68 | 100.00 | 176 |
| Thermotogae | 9 | 17,037 | 1.60 | 100.00 | 684 |
| α-proteobacteria | 94 | 347,287 | 8.66 | 100.00 | 121 |
| β-proteobacteria | 56 | 266,362 | 3.71 | 100.00 | 311 |
| γ-proteobacteria | 126 | 483,632 | 10.63 | 79.67 | 118 |
| δ-proteobacteria | 25 | 102,115 | 4.42 | 100.00 | 206 |
| ε-proteobacteria | 18 | 33,416 | 2.43 | 100.00 | 455 |

Table 2: Summary of the 40 PhyEco marker candidates identified for the group "Bacteria plus Archaea."

| Marker ID | Gene Family Descriptions | Correspondent AMPHORA Marker |
|---|---|---|
| BA00001 | ribosomal protein S2 | rpsB |
| BA00002 | ribosomal protein S10 | rpsJ |
| BA00003 | ribosomal protein L1 | rplA |
| BA00004 | translation elongation factor EF-2 | - |
| BA00005 | translation initiation factor IF-2 | - |
| BA00006 | metalloendopeptidase | - |
| BA00007 | ribosomal protein L22 | - |
| BA00008 | ffh signal recognition particle protein | - |
| BA00009 | ribosomal protein L4/L1e | rplD |
| BA00010 | ribosomal protein L2 | rplB |
| BA00011 | ribosomal protein S9 | rpsI |
| BA00012 | ribosomal protein L3 | rplC |
| BA00013 | phenylalanyl-tRNA synthetase beta subunit | - |
| BA00014 | ribosomal protein L14b/L23e | rplN |
| BA00015 | ribosomal protein S5 | - |
| BA00016 | ribosomal protein S19 | rpsS |
| BA00017 | ribosomal protein S7 | - |
| BA00018 | ribosomal protein L16/L10E | rplP |
| BA00019 | ribosomal protein S13 | rpsM |
| BA00020 | phenylalanyl-tRNA synthetase $\alpha$ subunit | - |
| BA00021 | ribosomal protein L15 | - |
| BA00022 | ribosomal protein L25/L23 | - |
| BA00023 | ribosomal protein L6 | rplF |
| BA00024 | ribosomal protein L11 | rplK |
| BA00025 | ribosomal protein L5 | rplE |
| BA00026 | ribosomal protein S12/S23 | - |
| BA00027 | ribosomal protein L29 | - |
| BA00028 | ribosomal protein S3 | rpsC |
| BA00029 | ribosomal protein S11 | rpsK |
| BA00030 | ribosomal protein L10 | - |
| BA00031 | ribosomal protein S8 | - |
| BA00032 | tRNA pseudouridine synthase B | - |
| BA00033 | ribosomal protein L18P/L5E | - |
| BA00034 | ribosomal protein S15P/S13e | - |
| BA00035 | Porphobilinogen deaminase | - |
| BA00036 | ribosomal protein S17 | - |
| BA00037 | ribosomal protein L13 | rplM |
| BA00038 | phosphoribosylformylglycinamidine cyclo-ligase | rpsE |
| BA00039 | ribonuclease HII | - |
| BA00040 | ribosomal protein L24 | - |

Table 1S: Summary of 74 PhyEco marker candidates identified for the group "Bacteria".

| Marker ID | Gene Family Descriptions | Correspondent AMPHRA-I Marker |
|---|---|---|
| B000041 | transcription elongation protein NusA | nusA |
| B000042 | rpoB DNA-directed RNA polymerase subunit beta | rpoB |
| B000043 | GTP-binding protein EngA | - |
| B000044 | rpoC DNA-directed RNA polymerase subunit beta' | - |
| B000045 | priA primosome assembly protein | - |
| B000046 | transcription-repair coupling factor | - |
| B000047 | CTP synthase | pyrG |
| B000048 | secY preprotein translocase subunit SecY | - |
| B000049 | GTP-binding protein Obg/CgtA | - |
| B000050 | DNA polymerase I | - |
| B000051 | rpsF 30S ribosomal protein S6 | - |
| B000052 | poA DNA-directed RNA polymerase subunit alpha | - |
| B000053 | peptide chain release factor 1 | - |
| B000054 | rplI 50S ribosomal protein L9 | - |
| B000055 | polyribonucleotide nucleotidyltransferase | - |
| B000056 | tsf elongation factor Ts | tsf |
| B000057 | rplQ 50S ribosomal protein L17 | - |
| B000058 | tRNA (guanine-N(1)-)-methyltransferase | rplS |
| B000059 | rplY probable 50S ribosomal protein L25 | - |
| B000060 | DNA repair protein RadA | - |
| B000061 | glucose-inhibited division protein A | - |
| B000062 | Unknown protein | - |
| B000063 | ribosome-binding factor A | - |
| B000064 | DNA mismatch repair protein MutL | - |
| B000065 | smpB SsrA-binding protein | smpB |
| B000066 | N-acetylglucosaminyl transferase | - |
| B000067 | S-adenosyl-methyltransferase MraW | - |
| B000068 | UDP-N-acetylmuramoylalanine--D-glutamate ligase | - |
| B000069 | rplS 50S ribosomal protein L19 | - |
| B000070 | rplT 50S ribosomal protein L20 | rplT |
| B000071 | ruvA holliday junction DNA helicase | - |
| B000072 | ruvB Holliday junction DNA helicase B | - |
| B000073 | serS seryl-tRNA synthetase | - |
| B000074 | rplU 50S ribosomal protein L21 | - |
| B000075 | rpsR 30S ribosomal protein S18 | - |
| B000076 | DNA mismatch repair protein MutS | - |
| B000077 | rpsT 30S ribosomal protein S20 | - |
| B000078 | DNA repair protein RecN | - |
| B000079 | frr ribosome recycling factor | frr |
| B000080 | recombination protein RecR | - |
| B000081 | protein of unknown function UPF0054 | - |
| B000082 | miaA tRNA isopentenyltransferase | - |
| B000083 | GTP-binding protein YchF | - |
| B000084 | chromosomal replication initiator protein DnaA | - |
| B000085 | dephospho-CoA kinase | - |
| B000086 | 16S rRNA processing protein RimM | - |
| B000087 | ATP-cone domain protein | - |
| B000088 | 1-deoxy-D-xylulose 5-phosphate reductoisomerase | - |
| B000089 | 2C-methyl-D-erythritol 2,4-cyclodiphosphate synthase | - |
| B000090 | fatty acid/phospholipid synthesis protein PlsX | - |
| B000091 | tRNA(Ile)-lysidine synthetase | - |
| B000092 | dnaG DNA primase | dnaG |
| B000093 | ruvC Holliday junction resolvase | - |
| B000094 | rpsP 30S ribosomal protein S16 | - |
| B000095 | Recombinase A recA | - |
| B000096 | riboflavin biosynthesis protein RibF | - |
| B000097 | glycyl-tRNA synthetase beta subunit | - |
| B000098 | trmU tRNA (5-methylaminomethyl-2-thiouridylate)-methyltransferase | - |
| B000099 | rpmI 50S ribosomal protein L35 | - |
| B000100 | hemE uroporphyrinogen decarboxylase | - |
| B000101 | Rod shape-determining protein | - |
| B000102 | rpmA 50S ribosomal protein L27 | rpmA |
| B000103 | peptidyl-tRNA hydrolase | - |
| B000104 | translation initiation factor IF-3 | infC |
| B000105 | UDP-N-acetylmuramyl-tripeptide synthetase | - |
| B000106 | rpmF 50S ribosomal protein L32 | - |
| B000107 | rplL 50S ribosomal protein L7/L12 | rplL |
| B000108 | leuS leucyl-tRNA synthetase | - |
| B000109 | ligA NAD-dependent DNA ligase | - |
| B000110 | cell division protein FtsA | - |
| B000111 | GTP-binding protein TypA | - |
| B000112 | ATP-dependent Clp protease, ATP-binding subunit ClpX | - |
| B000113 | DNA replication and repair protein RecF | - |
| B000114 | UDP-N-acetylenolpyruvoylglucosamine reductase | - |

Figures

Figure 1: Flow chart of the PhyEco marker identification pipeline.

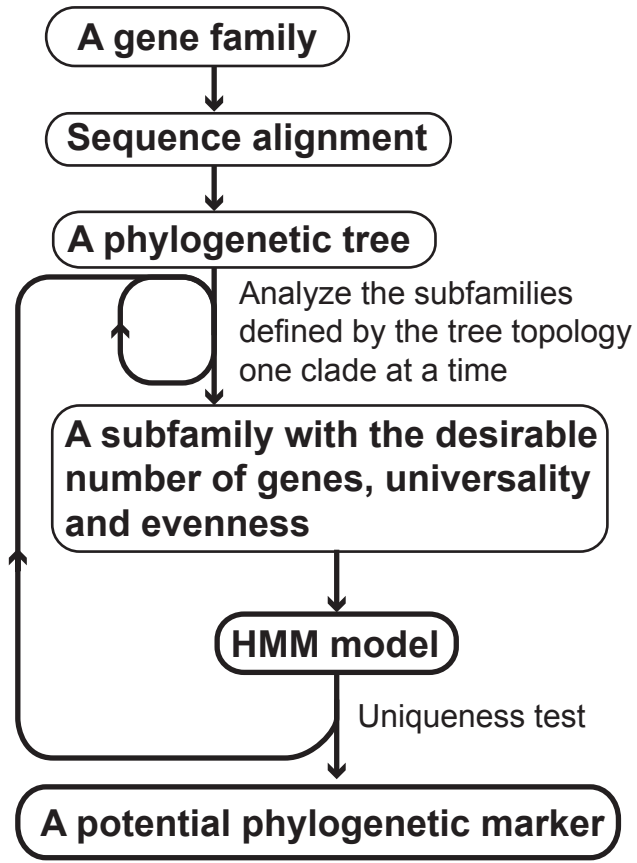

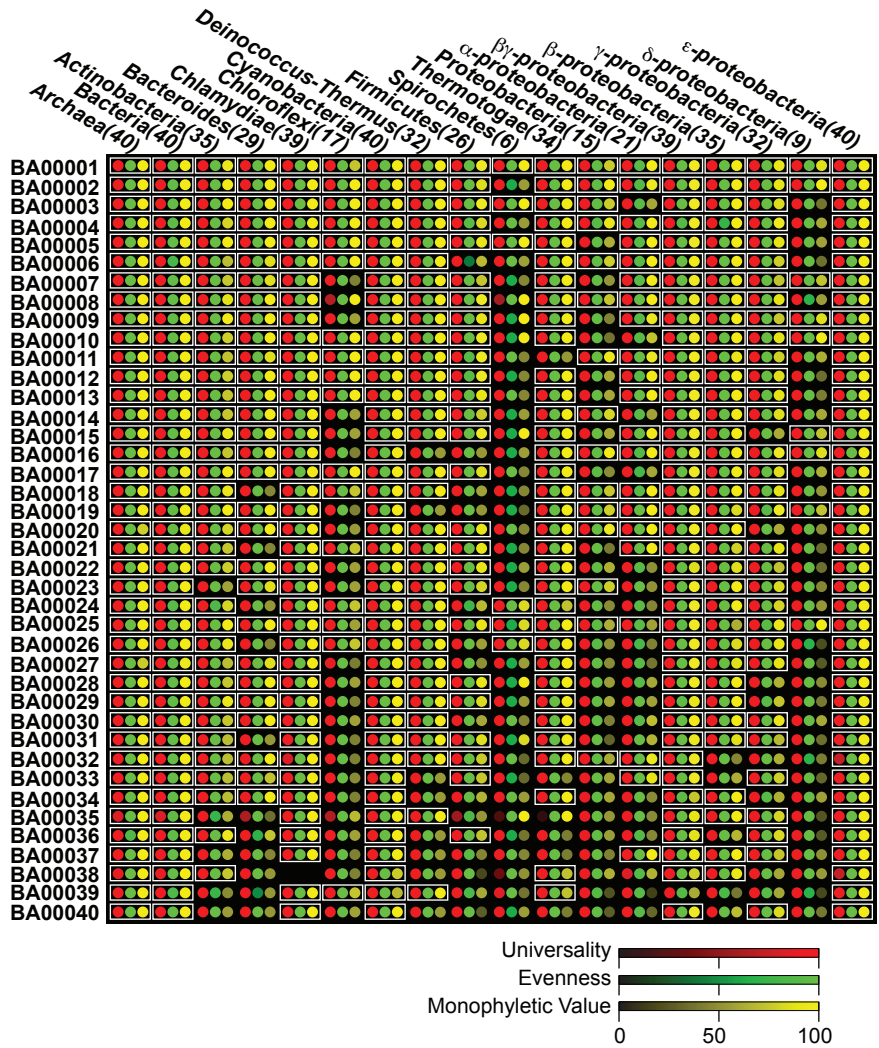

Figure 2: The universality, evenness and monophyletic value of the 40 Bacterial/Archaeal PhyEco marker candidates in different taxonomic groups. PhyEco marker genes for the taxonomic groups are highlighted with white boxes.

Figure 3: The universality, evenness and monophyletic value of the 74 Bacterial specific PhyEco marker candidates in different taxonomic groups. PhyEco marker genes for the taxonomic groups are highlighted with white boxes.

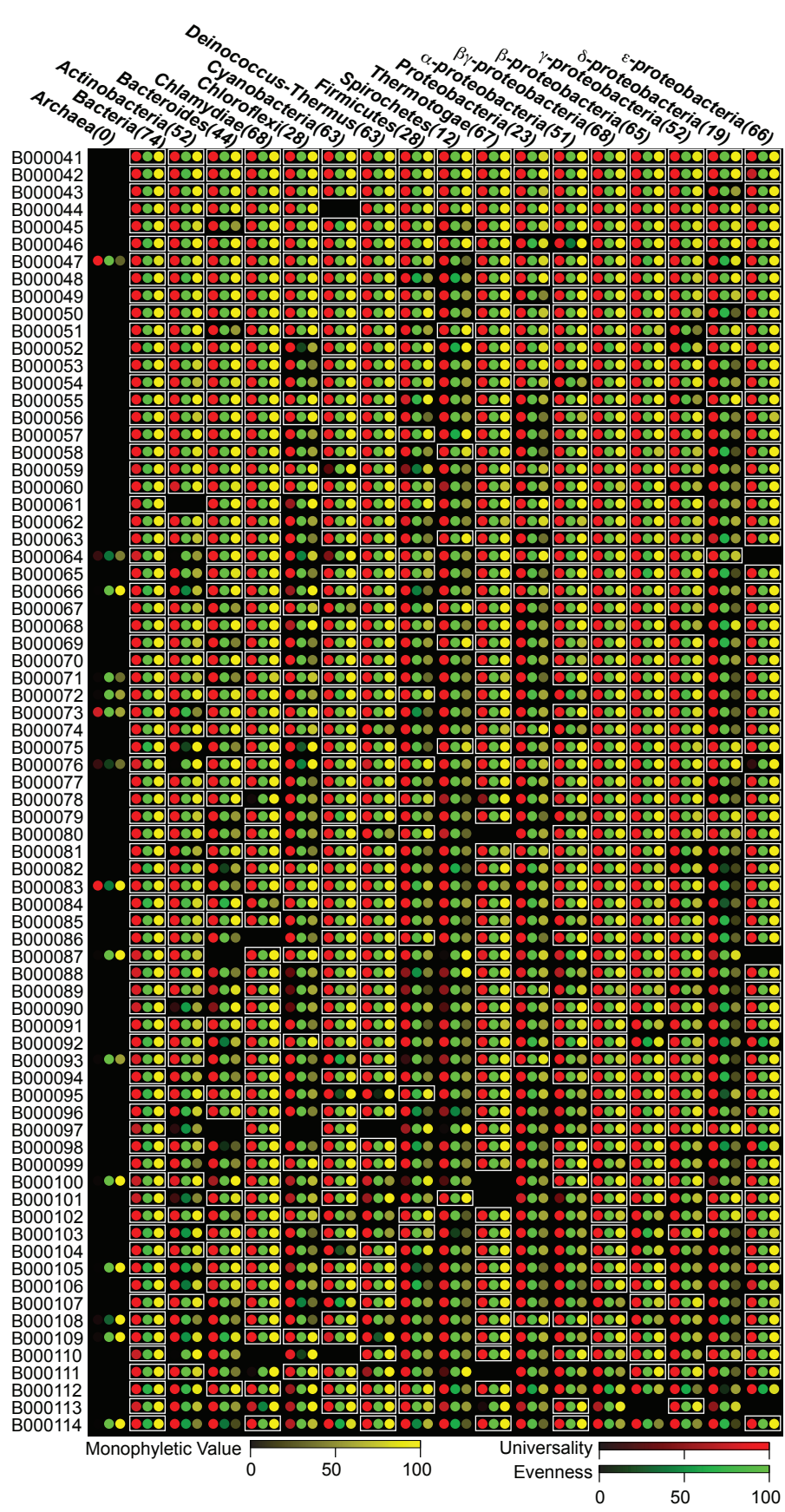